\documentclass[aps,reprint,nofootinbib,twocolumn,amsmath]{revtex4-1}
\usepackage{CJK}
\usepackage{graphicx,color}
\usepackage{lipsum}
\usepackage{bm}

\newcommand{\kk}{{\bf k}}
\newcommand{\hk}{{\hat k}}
\newcommand{\hz}{{\hat z}}

\newcommand{\uu}{{\bf u}}

\newcommand{\rr}{{\bf r}}
\newcommand{\KK}{{\widetilde K}}

\newcommand{\GG}{{\widetilde G}}
\newcommand{\MM}{{\widetilde M}}

\newcommand{\Gg}{{\sf G}}

\newcommand{\tr}{\mbox{Tr}}
\newcommand{\be}{\begin{equation}}
\newcommand{\ee}{\end{equation}}
\newcommand{\beq}{\begin{eqnarray}}
\newcommand{\eeq}{\end{eqnarray}}
\newcommand{\ba}{\begin{eqnarray}}
\newcommand{\ea}{\end{eqnarray}}
\newcommand{\bse}{\begin{subequations}}
\newcommand{\ese}{\end{subequations}}
\newcommand{\ds}{\displaystyle}
\newcommand{\im}{\mbox{Im}}
\newcommand{\re}{\mbox{Re}}
\newcommand{\ts}{\textstyle}

\begin{document}
		\title{Heterogeneous Elasticity: The tale of the boson peak}
		
		\author{Walter Schirmacher$^1$ and Giancarlo Ruocco$^{2,3}$}
		
		\affiliation{$^1$Institut f$\ddot{u}$r Physik, Universit$\ddot{a}$t Mainz, Staudinger Weg 7, D-55099 Mainz, Germany}
\affiliation{$^2$Fondazione Istituto Italiano di
Tecnologia (IIT), Center for Life Nano Science, Viale Regina Elena
291, I00161 Roma, Italy}
\affiliation{$^3$Department of
Physics, University of Rome `La Sapienza', Piazzale Aldo Moro, 5,
I00185, Rome, Italy}
\begin{abstract}
The vibrational anomalies of glasses, in particular the boson peak,
are addressed from the standpoint of heterogeneous elasticity, 
namely the spatial fluctuations of elastic constants caused by
the structural disorder of the amorphous materials.

In the first part of this review article
a mathematical analogy between
diffusive motion in a disordered environment and a scalar simplification
of vibrational motion
under the same condition is emploited.
We demonstrate that the
disorder-induced long-time tails of diffusion correspond to the
Rayleigh scattering law in the vibrational system and that the
cross-over from normal to anomalous diffusion corresponds to
the boson peak. The anomalous motion arises as soon as the disorder-induced
self-energy exceeds the frequency-independent diffusivity/elasticity.
For this model a variational scheme is emploited for deriving two mean-field
	theories of disorder, the self-consistent Born approximation (SCBA)
	and coherent-potential approximation (CPA).
	The former applies
	if the fluctuations are weak and Gaussian, the latter applies
	for stronger and non-Gaussian fluctuations. 

In the second part the vectorial theory of heterogenous elasticity
is presented and solved in SCBA and CPA, introduced for the scalar model.
	Both approaches
	predict and explain the boson-peak and the associated anomalies,
	namely a dip in the acoustic phase velocity and a characteristic
	strong increase of the acoustic attenuation below the boson peak.
	Explicit expressions for the density of states and the
	inelastic Raman, neutron and X-ray scattering laws are given.
	Recent conflicting ways of explaining the boson-peak anomalies
	are discussed.

\end{abstract}

\maketitle

\section{Introduction}
The vibrational properties of disordered solids are quite different
from those of crystals \cite{elliott,kobbinder11}. 
While the harmonic vibrational spectra of crystals are given by
the dispersion relations obtained from the crystalline symmetry groups,
those of amorphous solids, in particular glasses exhibit anomalous,
continuous spectra, which are a matter of debate since 60 years 
\cite{comment1,nakayama02,klinger10,kobbinder11,schirm13,schirm14}.

The first evidence of something unusual in the vibrational spectra of
glasses came from Raman scattering
\cite{flubacher59,leadbetter63,leadbetter68,leadbetter69,shuker70,shuker71}.
In the THz or 100 wavenumber regime one observed a broad maximum,
where usually either no intensity or very sharp peaks due to 
low-lying optical
modes were observed. Because this maximum obeyed the frequency/temperature
dependence of the Bose function $n(\omega)+1=[1-exp\{-\hbar\omega/k_BT\}]^{-1}$
($\omega/2\pi$ is the frequency and $k_BT$ is the Boltzmann constant
times the temperature)
one called this maximum
``boson peak''  \cite{jackle81}. Here we note that because the Raman
scattering intensity is proportional to $n(\omega)+1$ times the
Raman spectral function $\chi_R''(\omega)$ \cite{shuker71,martinbrenig,schmid09}, the latter must be temperature independent if the entire temperature
dependence comes from the Bose function. This points to a harmonic
origin of the boson peak and questions all interpretations in terms
of anharmonic interactions.

A maximum in the frequency range $\sim$ 5 meV $\sim$ 1 THz
was observed in inelastic coherent neutron scattering data
of several glasses
\cite{buchenau84}.
This maximum appeared to be related to an excess of the vibrational density
of states (DOS) $g(\omega)$ with respect to Debye's $g(\omega)\propto \omega^2$
law, which appears as a peak in the reduced DOS $g(\omega)/\omega^2$.
This was confirmed experimentally by means of inelastic 
incoherent neutron \cite{wuttke95} and nuclear scattering
\cite{chumakov04}, which both measures directly the DOS.
Further, the boson peak turned out to be related to a maximum in the
``reduced'' specific heat $C(T)/T^3$ in the 10K regime \cite{zellerpohl,wuttke95}, which is also called ``boson peak''. Right at the boson-peak temperature the thermal conductivity of the same glasses exhibits a characteristic shoulder or dip
in its temperature dependence
\cite{freeman86}, which turned out to be an ``upside-down boson peak''
\cite{schirm06}.

These are not the only low-temperature thermal anomalies of glasses.
Below the boson peak, in the $\sim$ 1K regime, the specific heat does not show
Debye's $T^3$ law, but varies approximately linearly with temperature.
The thermal conductivity varies quadratically in this regime. These
findings have been attributed to the existence of bistable structural
arrangements, which allow for tunneling between the two positions, giving
rise to a tunnel splitting (two-level systems (TLS), tunneling systems) 
\cite{phillips72,anderson72,phillips87,yuleggett88,wuerger96}. If the energy separations
of the TLS are assumed to have a broad distribution, the 1K anomalies
can be explained. Independent evidence for the existence of the TLS
comes from ultrasonic and nuclear magnetic resonance data 
\cite{hunklinger86,hunklinger97}.

Inelastic X-ray scattering contributed more anomalous
features in the boson-peak regime \cite{sette98}. In particular
the group velocity of longitudinal sound (dispersion of the Brillouin
line position)
was found to exhibit a minimum
at the wavenumber corresponding to the boson peak, and the sound attenuation
(width of the Brillouin line) 
was found to increase strongly with frequency
in the boson-peak regime (Rayleigh scattering) \cite{monaco09,baldi09,baldi10,ruta10,baldi11,ruta12}. These anomalies were confirmed in computer simulations
\cite{monacomossa09,marruzzo13}.
We call all these features the boson-peak related anomalies.

The theoretical interpretation of the boson peak has a very diverse history
and gave rise to several controversies \cite{comment1}.
In one line of argumentation for explaining the boson-peak related anomalies 
the authors tried to find a classical analogon of the tunneling model, the
soft-potential model \cite{karpov83,buchenau91,comment2}. The bistable configurations,
which can be described by an anharmonic
double-well potential with a rather shallow
barrier between the wells, were assumed to have a distribution of their
characteristic parameters, in particular the second-order coefficient.
It was found that in general the statistics of the vibrational
excitations of such defect potentials would give rise to a constant
density of eigenvalues $g(\omega^2)$, which leads to a DOS
$g(\omega)\propto \omega$. On the other hand, so the argument,
at lower frequencies (below a cross-over frequency $\omega_c$)
the balance between unstable configurations
with negative second-order coefficients and
the anharmonic fourth-order ones give rise to a defect DOS
$\propto \omega^4$. Therefore the
reduced DOS $g(\omega)/\omega$ increases $\propto \omega^2$
below $\omega_c$ and decreases $\propto\omega^{-1}$ above
$\omega_c$. In between, i.e. near $\omega_c$ is then the boson peak
\cite{karpov83,schober14}. 

Arguments that the boson peak, by its temperature characteristics,
is not an
anharmonic phenomenon were met by
the authors of the soft-potential model by noting
that the anharmonic
interaction only acts in producing the soft configurations
in the quenching process \cite{gurevich03}.
In the quenched state they are 
supposed to act like local 
harmonic oscillators,
similar to heavy-mass atoms, coupled to the acoustic
harmonic degrees of freedom \cite{eco71,burin95,maurer04,schirm11}.

Another attempt to explain the boson-peak anomaly was the phonon-fracton
model \cite{ao82,derrida84,entin84,alexander86,nyo94}. It was
postulated that disordered solids exhibit a certain degree of
fractal structure. A fractal is a self-similar structure
\cite{mandelbrot,schirmacher15}, which has a non-integer
dimensionality $D_0$, which is smaller than the embedding
dimensionality $d$. Real fractals like sponges or trees
have a smallest and largest scale, in which the self-similarity
holds. The smallest scale is e.g. the smallest pore diameter
of a sponge, the largest is the correlation length $\xi$. For
scales larger than $\xi$ the object looks like an ordinary
material in which the mass scales as $L^d$, where $L$ is the size.
For scales smaller than $\xi$ the mass scales as 
$L^{D_0}$.
Alexander and Orbach \cite{ao82} have shown that the vibrational
degrees of freedom scale with a fractal dimensionality $d_s<D_0$
(spectral dimensionality) and that the DOS of such an object
obeyes a Debye law below $\omega_\xi=2\pi v/\xi$, where $v$ is
the sound velocity. Above $\omega_\xi$ the DOS behaves as
$g(\omega)\propto \omega^{d_s-1}$. They found that in all
fractal structures they investigated $d_s\approx 4/3$.
The specific model employed by the phonon-fracton supporters
was a percolating lattice, i.e. a cubic lattice in which
a certain percentage of bonds
$1-p$ 
(carrying neares-neighbor force constants)
was missing. If the bond concentration
$p$ is larger than the critical concentration $p_c$, which determines
the connectedness of the structure a finite correlation length
$\xi\propto( p-p_c)^\nu$ exists ($\nu$ is the order-parameter 
exponent \cite{nyo94}). Calculations using the coherent-potential
approximation (CPA) \cite{cpa1,cpa2,cpa} showed that in between
the Debye and the fracton regime an enhancement over the
Debye $g(\omega)\propto \omega^2$ law was present \cite{derrida84,entin84}. 
This was, for the time being, a
satisfactory explanation of the boson peak.
However, numerical simulations of the percolation-phonon-fracton model
\cite{nyo94} 
showed that the phonon-fracton crossover in the
DOS of this model occurs very smooth without any excess over
the Debye law. Obviously the excess in the calculations 
\cite{derrida84,entin84}
had been an artifact of the
CPA. Another argument against the phonon-fracton model 
as candidate for explaining the boson-peak anomalies
is that - apart from
aerogels \cite{courtens87,caponi09} 
glasses do not show any self-similary, which
should show up (but does not)
as an enhanced small-angle scattering in neutron or X-ray
diffraction data.

In other articles reflecting on the vibrational anomalies of glasses
the boson peak and the anomalous
shoulder in the temperature dependence of the thermal conductivity
were also considered to be related to Anderson localization
of sound \cite{am85,alexander86,graebner86}. It was observed that
near the boson-peak frequency the mean-free path of the acoustic
excitations is of the order of the sound wavelength. According
to a rule, coined by Ioffe and Regel in their survey of electronic
conduction in semiconductors
\cite{iofferegel60}, the notion
of a mean-free path, which implies a wave, which is occasionally
scattered by an inhomogeneity, breaks down once the mean-free path
becomes equal to the wavelength (Ioffe-Regel limit). 
Mott \cite{Mott90} conjectured that Anderson localization occurs
for electrons
near the Ioffe-Regel energy. For phonon it was assumed \cite{akkermans85,graebner86}, that
the crossover from extended to Anderson-localized states would take place near
the Ioffe-Regel frequency and that the boson peak would mark the onset
of localized states. In particular it was thought, that the presence of
localized states would cause the dip in the thermal conductivity
\cite{graebner86}.
Later, more detailed theoretical
investigations showed, however, that the Anderson transition
in realistic solids does not occur near the Ioffe-Regel crossover,
but in a much higher frequency range near the Debye frequency
\cite{john83,allen89a,schirm93,schirm98,pinski12,pinski12a,tomaras13}.

Later many researchers were intested in whether the boson-peak frequency
coincides exactly with the Ioffe-Regel frequency, given by the
implicit relation
$\omega_{IR}=\pi\Gamma(\omega_{IR})$, where $\Gamma(\omega)$ is the
sound attenuation coefficient
\cite{taraskin99,ruffle06}. $\Gamma(\omega)$ and hence $\omega_{IR}$
is, of course, different for longitudinal and transverse sound waves.
It appeared that in materials governed by hard-sphere-like potentials
the boson-peak coincides with the transverse Ioffe-Regel frequency
\cite{shintani08,marruzzo13}, whereas in network glasses with the
longitudinal one \cite{foret96,foret97,foret99,taraskin00}. 

In 1991 Schirmacher and Wagener \cite{schirm92} exploited the mathematical
analogy between a single-particle random walk and harmonic phonons
\cite{alexander81} using an off-lattice version of the CPA.
They demonstrated that the cross-over from
a frequency-independent conductivity/diffusivity to a frequency-dependent
one \cite{jonscher77,long82} corresponds to the onset of the frequency
dependence of the complex sound velocity. The latter was found to lead
to a boson peak in the vibrational DOS. 

Because understanding this analogy is essential for grasping the
essence of the BP-related anomalies,
we devote a whole section
of the present review to this analogy.

However, because the CPA in the case of the percolating
lattice predicted a boson peak \cite{derrida84,entin84}, which did not exist in
the simulation of the same system \cite{nyo94}
a check was needed, whether the resulting
boson-peak enhancement of the DOS
was not an artifact of the CPA like in the phonon-fracton
model. Therefore in Ref. \cite{schirm98}
the lattice version of the CPA \cite{webman81,odagaki81,summerfield81}
was compared to a numerical calculation for a cubic lattice with
fluctuating nearest-neighbor force constants. Both simulation
and CPA calculation showed a boson peak, and good agreement between
CPA and the numerical spectrum was found. So it was demonstrated that
effective-medium calculations are reliable for investigating the 
influence of disorder on the harmonic spectrum of a model solid.
The breakdown of the CPA in the case of the phonon-fracton model
was obviously due to the critical fluctuations in this model,
which are not generic for disordered solids.

The boson peak in the model calculations of Ref. \cite{schirm98}
was identified to be caused by very small positive and negative force
constants, and being a precursor of an instability, which happens
for stronger disorder \cite{pinski12,pinski12a}.

With the help of the numerical calculation in Ref. \cite{schirm98}
it could also decided, whether the vibrational states near and
above the boson peak were localized or extended. This was
achieved by means of the statistics of the distances between
the eigenvalues. Near and above the boson peak they showed the
so-called Gaussian-ortogonal-ensemble (GOE) statistics
of random-matrix theory \cite{mehta67,izrailev90}, which proves
that the corresponding states are delocalized. At high frequencies,
near the upper band edghe a transition to Poissonian statistics
was observed, which is evidence for a delocalization-localization
transition in this regime, in agreement with earlier
estimates \cite{john83,allen89a}. On the other hand, the GOE statistics
is a sign for the so-called level repulsion, showing that each eigenvalue
is non-degenerate due to the absence of symmetries in the disordered
system. As the generic spectrum of random matrices is not a Debye
spectrum (which is highly degenerate) the boson peak marks the transition
from a Debye to a random-matrix-type spectrum \cite{schirm13}.

The spectra calculated in Ref. \cite{schirm98}
for a cubic lattice with very small disorder exhibited
the usual van-Hove singularities, which appear as a result of the
leveling-off of the crystalline phonon dispersions $\omega(k)$
at the Brillouin-zone boundary \cite{ashcroft76}. 
With increasing
disorder the data showed that
the sharp van-Hove peak became rounder and is shifted
downwards and gradually transformed to the low-frequency boson peak.
This led Taraskin et al. \cite{te01} to the conclusion that
the boson peak is just a crystal-like van-Hove peak, modified by disorder.
This appeared as a rather unexpected conclusion, because 
a van-Hove singularity is a typical signature of a crystalline
structure with long-range order and was not known to exist in
glasses. However, until now, the boson peak explanation as
a glassy version of a van-Hove singularity is still considered
to be an alternative to the disorder explanation
\cite{chumakov11,zorn11,chumakov14}. We further comment on this
in section IV.

Quite recently in an
experimental study of
a macroscopic disordered model glass it was shown that the disorder-induced
maximum of the reduced DOS and that induced by the van-Hove mechanism
are, in fact, two different phenomena \cite{wang18}. 

The random-matrix aspect of the disorder-vibration problem was elaborated
further in the literature
\cite{matharoo04,kuhn97,mezard99,parisi00,parisi01,parisi01a,parisi03a,%
parisi03b,ganter11,parisi11,beltukov13}, 
in particular by means of the
euclidean random-matrix theory
\cite{mezard99,parisi00,parisi01,parisi01a,parisi03a,%
parisi03b,ganter11,parisi11}.

As in Ref. \cite{schirm98} a disorder-induced boson peak was found and
shown to be the precursor of an instability (``phonon-saddle
transition'' \cite{parisi03a}).

A quite different and interesting approach to the boson-peak anomaly
was worked out \cite{gotzemayr99} in the framework of the mode-coupling
theory of the glass transition . This theory
of glassy freezing in its original form \cite{beng_kinhs,gotze}
describes the idealized glass transition as a dynamical transition
towards a non-ergodic state leading to a frozen-in additional contribution
to the static longitudinal susceptibility. This, in turn leads to
a characteristic hump in the density fluctuation spectrum
of the idealized glass
(proportional to the neuton scattering law $S(k,\omega)$), which
was identified with the boson-peak anomaly found in neutron
scattering experiments. Interestingly this theory already
predicted the characteristic
dip in the longitudinal group velocity,
which was found later to be associated with the boson peak, as
mentioned above
\cite{monaco09,baldi11,%
monacomossa09,marruzzo13}.

We feel that an important step in the understanding of the boson peak was
achieved by working out heterogeneous-elasticity theory \cite{schirm06}.
In this phenomenological theory it is assumed that the shear modulus
in ordinary elasticity theory \cite{landau59} is assumed to exhibit
spatial fluctuations. The resulting stochastic equations were solved
by field-theoretical techniques \cite{mckane81,john83}, resulting
in a mean-field theory, called self-consistent Born approximation
(SCBA). The SCBA is obtained from assuming a Gaussian distribution
of the elasticity fluctuations and that the relative width of the
Gaussian is a small parameter (``disorder parameter'' $\gamma$).
Again, if the disorder parameter becomes larger than a critical
one, an instability occurs, which is due to too many regions with
negative elastic constants.

Shortly before heterogeneous-elasticity theory had been worked out
a series of papers appeared, in which molecular-dynamics simulations
of glasses were investigated for their elastic and vibrational properties
\cite{wittmer02,leonforte05,leonforte06,tsamados09}. It turned out
that, by applying external forces, indeed heterogeneous shear deformations
are present. The authors showed that in regions with strong deformations
the shear response is
highly non-affine, i.e. the displacements do not
follow the direction of the applied stress. In the non-affine regions
the local shear moduli were found to be very small and even negative,
a finding, which was observed also in other simulations
\cite{mayr09,mizuno13,marruzzo13}. This nicely confirmed the model
assumption of heterogeneous-elasticith theory. In Ref. \cite{marruzzo13}
a direct comparison between the elasticity fluctuations in a simulated
glass and the theory of Ref. \cite{schirm06} was made and good agreement
was found. We shall comment on this article in more detail below.

In later simulations it was shown \cite{zaccone16d,zaccone19d} that the
regions in the simulated disordered solids, which have pronounced 
non-affine response, are also regions with strong inversion-symmetry
breaking. The authors introduced quantitative measures of
this inversion-symmetry breaking and found a unique correlation
with the height of the boson peak. This was also found
in the quoted investigation of
a macroscopic model of a disordered solid, in which 
soft spots with strong
inversion-symmetry breaking were shown to contribute predominantly
to the boson peak \cite{wang18}.

An important aspect of the boson peak appeared when it was
shown \cite{schirm07} with the help of heterogeneous-elasticity theory
that the excess DOS 
$\Delta g(\omega)=g(\omega)-g_D(\omega)$ with respect to the
Debye DOS $g_D(\omega)$ is proportional to the sound attenuation
in the boson-peak frequency range. Heterogeneous-elasticity
provides an expression for the disorder-induced sound attenuation
as imaginary part of frequency-dependent elastic coefficients
(see section ..). These enter into the spectral functions,
and the DOS enhancement is just produced by the sound attenuation,
which increases rapidly in the boson-peak frequency regime
due to Rayleigh scattering.

The rest of this contribution is organized as follows:
In section II. the mathematical analogy between diffusion and
scalar elasticity is investigated in detail and two mean-field
or effective-medium theories, the self-consistent Born approximation
(SCBA) and the coherent-potential approximation (CPA) for this
model are introduced and solved. A pedagogical derivation, which
is more simple than the original field-theoretical one, is presented.
We show that the boson-peak related anomalies are the analogon of
the crossover from a frequency-independent diffusivity/conductivity 
to a
frequency-dependent one.

In section III. the full vectorial heterogeneous elasticity
theory is presented and solved in SCBA and CPA. The salient
features of the boson-peak related anomalies of glasses are
discussed with the help of these mean-field theories.
In section IV. we discuss recent conflicting theories
of these anomalies.

\section{``Scalar elasticity'' and Diffusion-vibration analogy}
\subsection{Diffusion-vibration analogy}
A simplified version of heterogeneous elasticity theory, which proved
to be helpful in understanding the spectral properties of disordered
solids \cite{schirm92,parisi03a,maurer04,maurer04a,ganter10,kohler13}
is represented by a scalar wave equation with a spatially
fluctuating elastic constant $K(\rr)\equiv \rho\KK(\rr)$
($\rho$ is the mass density)
\be\label{scalart}
\frac{\partial^2}{\partial t^2}u(\rr,t)=\nabla \KK(\rr)\nabla u(\rr,t)
\ee
The elastic coefficient $\KK(\rr)$, which is the square
of the local sound velocity, $v(\rr)$ may be sub-divided into
an average elasticity $\KK_0=\langle \KK \rangle$ and deviations
$\Delta \KK(\rr)$
\be
\KK(\rr)=v^2(\rr)=\KK_0+\Delta \KK(\rr)
\ee
If we replace the second time derivative in Eq. (\ref{scalart})
by a first one, we arrive at a diffusion equation, which describes
the random walk of a particle, which encounters a spatially varying
diffusivity, e.g. due to a spatially varying activation energy
\ba\label{scalard}
\frac{\partial}{\partial t}n(\rr,t)&=&\nabla D(\rr)\nabla n(\rr,t)\nonumber\\
&=&\nabla \bigg[D_0+\Delta D(\rr)\bigg]\nabla n(\rr,t)
\ea
with $D_0$, again, denoting the avarage diffusivity and
$\Delta D(\rr)$ the fluctations.
$n(\rr,t)$ is the probability density for finding the particle
within a volume element around $\rr$ at time $t$. 
The Green's function of Eq. (\ref{scalart}) in frequency space obeys
the equation
\ba\label{green1a}
&&\bigg(-z^2-\nabla \KK(\rr)\nabla \bigg)G(\rr,\rr')\nonumber\\
&=&\bigg(-z^2-\nabla \big[\KK_0+\Delta\KK(\rr)\big]\nabla \bigg)G(\rr,\rr')\nonumber\\
&\equiv&A[z,\rr,\KK(\rr)]G(\rr,\rr')=
\delta(\rr-\rr')
\ea
with $z=\omega+i\epsilon, \epsilon\rightarrow +0$. 
The operator $A[z,\rr,\KK(\rr)]$ is the operator-inverse of
the Green's function.
On the other
hand the Green's function corresponding to the heterogeneous diffusion
equation (\ref{scalard}) in frequency space is
\be\label{green1b}
s-\nabla D(\rr)\nabla G(\rr,\rr')=\delta(\rr-\rr')
\ee
with $s=i\omega+\epsilon$.

{\it So all calculations done for the scalar vibration problem (\ref{green1a})
can be taken over for the diffusion problem (\ref{green1b}) provided
we identify 
$D\leftrightarrow \KK$, 
$s\leftrightarrow -z^2$, or
~~$i\omega\leftrightarrow -\omega^2$.} 

In the next three subsection we shall demonstrate 
by means of different approximation schemes
(Born approx, Self-consistent Born approx. and
coherent-potential approx.)
that the
quenched glassy disorder induces a characteristic frequency
dependence to the macroscopic diffusivity/elasticity.
So these approximation schemes act as a coarse-graining
scheme, which converts spatial fluctuations to frequency
dependences.

In this context we may distinguish between three characteristic
scales: the microscopic scale, the mesoscopic scale and
the macroscopic one. The microscopic scale is the molecular
one and may be described by microscopic quantum or classical
equations of motions. The mesoscopic scale is a scale 
of 5 or 6 atomic or molecular diameter. This is the minimal
scale at which one may define local diffusivities or
elastic constants \cite{lutsko88,mayr09,kohler13,marruzzo13,mizuno13},
which exhibit spatial
fluctuations in structurally disordered materials. The macroscopic
scale is the experimental one (mm or cm), in which the 
macroscopic diffusivities
or elastic coefficients are frequency-dependent.

\subsection{Low frequency limit: Born approximation, Rayleigh
scattering and long-time tails}

The solution of Eq. (\ref{green1a}) or (\ref{green1b})
without fluctuations is given in $\kk$ space by
($\kk$ is the wave vector corresponding to $\rr-\rr'\equiv \tilde \rr$)
\be\label{green0}
G_0(k,z)=\frac{1}{-z^2+\KK_0k^2}
\ee
The disorder-averaged full Green's function should also only depend on $|\tilde\rr|$ only
and therefore may be represented as \cite{ganter10}
\ba\label{green1}
\langle G(z)\rangle_\kk&=&\Gg(\kk,z)=\frac{1}{-z^2+k^2(\KK_0-\Sigma(z))}\nonumber\\
&\equiv&\frac{1}{-z^2+k^2Q(z)}
\ea
where $\Sigma(z)=\Sigma'(\omega)+i\Sigma''(\omega)$
is the self-energy function, which
describes
the influence of the fluctuations $\Delta \KK(\rr)$
or $\Delta D(\rr)$.
Here we have defined 
a frequency-dependent 
elasticity
$Q(z)=\KK_0-\Sigma(z)$, corresponding to a frequency-dependent
diffusivity
$D(s)=D_0-\Sigma(s)$.
The former may be identified with the square
of a frequency-dependent sound velocity $v(z)$, i.e.
$Q(z)=v(z)^2$.

To lowest
order in the fluctuations one obtains by straightforward
perturbation theory \cite{ganter10} the Born approximation
\be\label{born}
\Sigma(z)=\gamma\frac{\ts 1}{\ts V}\sum_\kk k^2 G_0(\kk,z)\, ,
\ee
where
\mbox{$\sum_\kk\equiv\frac{V}{(2\pi)^3}\int d^3\kk$}, 
$V$ is the sample volume, and
\be
\gamma=
\langle\Delta \KK^2\rangle V_c\, ,
\ee
i.e. the variance of $\KK(\rr)$ times a coarse-graining
volume $V_c$, which serves to calculate the local elastic
coefficient \cite{lutsko88,ganter10,marruzzo13}.

If one imposes an upper cutoff $k_{\rm max}$
in the wavenumber integration, the integral in Eq. (\ref{born})
can be done exactly. For small frequencies we obtain
\be
\Delta\Sigma(s)=\Sigma(z)-\Sigma(0)\propto s^{3/2}
\ee
For the vibrational problem $s^{3/2}\rightarrow i\omega^3$.

We now show that this leads to 
Rayleigh's $\omega^4$ scattering law \cite{rayleigh71,rayleigh99}:

We may be interested in the wave intensity given by the
modulus of Eq. (\ref{green1})
\be\label{intensity}
|G(\tilde\rr,z)|^2=\left(\frac{1}{4\pi \KK_0}\right)^2
e^{\ds -\tilde r/\ell(\omega)}
\ee
with the mean-free path
\ba\label{rayleigh2}
\frac{1}{\ell(\omega)}&=&2\im\big\{\frac{\omega}{v(\omega)}\big\}\nonumber\\
&\approx&\frac{\Sigma''(\omega)\omega}{v_0^3}
=\frac{\gamma}{12\pi}\left(\frac{\omega}{v_0}\right)^4
\ea
where $v_0=\sqrt{\KK_0}$.

Eq. (\ref{rayleigh2}) constitutes the {\it Rayleigh scattering law}
\cite{rayleigh71,rayleigh99}. It holds for harmonic excitations
in the presence of quenched disorder \cite{ganter10,parisi11},
provided the disorder fluctuations do not exhibit long-range order
\cite{ganter10,john83a,gelin16}. 

Rayleigh scattering in Glasses is usually
obscured by anharmonic sound attenuation, which prevails in the sub-THz
frequency range. It has been observed experimentally in the THz regime in
some glasses
\cite{monaco09,baldi11} as well in computer simulations
\cite{monacomossa09,marruzzo13}.

In the diffusion problem $D(s)$ is
the frequency-dependent diffusivity, which
can be shown \cite{hansen06} to be the Laplace transform of
the velocity autocorrelation function $Z(t)$ of the moving particle
$D(s)=\int_0^\infty dt e^{-st}Z(t)$

We apply the Tauberian theorem \cite{feller71}
\be
\lim_{s\rightarrow 0}D(s)\propto s^{-\rho}\quad\Leftrightarrow\quad
\lim_{t\rightarrow\infty}D(t)\propto t^{\rho-1}
\ee
from which we conclude
\be
\lim_{t\rightarrow\infty}Z(t)\propto t^{-5/2}\, ,
\ee
a behaviour well known for particles performing a random walk
in a quenched-disordered environment \cite{ernst84,machta84}.

On the other hand,
by the Nernst-Einstein relation
\be\label{ne}
\sigma(s)=\sigma'(\omega)+i\sigma''(\omega)
=\frac{ne^2}{k_BT}
\ee
the frequency-dependent diffusivity is related to the dynamic
conductivity, the real part of which, $\sigma'(\omega)$ is the
alternate-current (AC) conductivity. 
Therefore, one expects a non-analytic low-frequency dependence
of the AC conductivity increment $\sigma'(\omega)-\sigma(0)
\propto \omega^{3/2}$. Indeed, such a behavior has been observed
in amorphous semiconductors \cite{long88}. We come back to this
in the subsection on strong disorder.
\subsection{%
Weak disorder and the
self-consistent Born approximation (SCBA)}
A well-known characteristic of the AC conductivity in disordered
materials is that beyond a characteristic frequency $\omega^*$
it starts to increase with frequency, in many cases with a 
characteristic power law $\sigma'(\omega)\propto \omega^\alpha$
where $\alpha$ is smaller than 1 and takes values
around 0.8 
\cite{jonscher77,%
long82,dyre00}.

For the random walker in the disordered environment
this means that the mean-square distance walked does
not increase linearly with time but sublinearly with
exponent $1-\alpha$. Such a behavior has been termed
anomalous diffusion. So the cross-over at $\omega^*$ 
corresponds to a transition from anomalous diffusion
for times $t<1/\omega^*$ to normal diffusion for $t>1/\omega^*$.

As pointed out in Refs. \cite{schirm92,kohler13} the cross-over
at $\omega^*$ - if transformed from the diffusion to the
scalar-vibrational system - corresponds to the boson peak.
In other words: it correspondss to the begin of the frequency
dependence of $\KK(\omega)=v(\omega^2)$. By the 
Kramers-Kronig
correspondence this implies the onset of an imaginary part
of $\KK(\omega)$ which becomes of the order of its real part.

A minimal theory for the boson peak, in fact, can be obtained
by the {\it self-consistent} version of the Born approxmation.
It is obtained by replacing the bare Green's function in 
Eq. (\ref{born}) by the full Green's function:
\be\label{scba0}
\Sigma(z)=\gamma\frac{\ts 1}{\ts V}\sum_{|\kk|\le k_{\rm max}} k^2 \Gg(\kk,z)
\ee
with $\Gg(k,z)$ given by Eq. (\ref{green1}). 
For the ultraviolet cutoff
$k_{\rm max}$ one should take the inverse of the length
scale which is the diameter of the coarse-graining volume $V_c$ 
used to define the local elastic constant/local diffusivity
\cite{ganter10}. On the other hand this length scale should
be of the order of the correlation length 
$\xi$ of the elasticity/diffusivity
fluctuations \cite{john83a}. In this work we treat these two
length scale as being the same. We therefore call the
cutoff $k_{\rm max}=k_\xi$.

While the ``derivation'' of Eq. (\ref{scba0}) is,
of course, just an ad-hoc replacement, 
a proper derivation is achieved
by field-theoretical techniques \cite{maurer04}
in analogy to the derivation
of the nonlinear sigma model for electrons 
\cite{wegner79,efetov80,schafer80,mckane81}, classical sound
waves \cite{john83,john83a} and electromagnetic waves
\cite{john87,schirmacher18}.

A royal road for this derivation is to minimize the following 
(dimensionless, frequency-dependent) mean-field free energy
or effective action with respect to $\Sigma(z)$
\be\label{seff}
S_{\rm eff}[\Sigma(z)]=S_{\rm med}[\Sigma(z)]
+S_{\rm SCBA}[\Sigma(z)]
\ee
with
\ba\label{seff1}
&&S_{\rm med}[\Sigma(z)]=S_{\rm med}[Q(z)]\\
&&=
\tr\ln\bigg(A\big[z,\rr,\underbrace{\KK_0-\Sigma(z)}_{\ds Q(z)}\big]\bigg)
=\sum_\kk \ln\bigg(A\big[z,\kk,Q(z)\big]\bigg)\nonumber
\ea
and
\be\label{seff2}
S_{\rm SCBA}[\Sigma(z)]=
\frac{1}{2}\frac{V}{\gamma}\Sigma(z)^2
\ee
The first term, 
$S_{\rm med}[\Sigma(z)]$, is the generalized
free energy of the effective medium \cite{vollhardt10}, which is
a medium without disorder, in which the
fluctuating force constants in the operator
$A[z,\rr,\KK(\rr)]$
are replaced by the
homogeneous (but frequency-dependent) force constant
$Q(z)=K_0-\Sigma(z)$. The trace can therefore
be calculated in $\kk$ space.
The Fourier transform of $A[z,\rr,Q(z)]$
is given by
\be
A[z,\kk,Q(z)]=-z^2+k^2Q(z)
\ee
This is no more the operator-inverse but just the ordinary
inverse of the mean-field Green's function
\ba
G(\kk,z)&=&
\frac{1}{A[z,\kk,Q(z)]}\nonumber\\
&=&\frac{1}{-z^2+k^2Q(z)}
\ea

The second term,
$S_{\rm SCBA}[\Sigma(z)]$
arises \cite{schirm06,john83}
from a Gaussian configuration
average of the full Green's function with distribution density
\be\label{gauss}
P\big[\Delta \KK(\rr)\big]=P_0\exp\{-\frac{1}{2\gamma}
\int d^3\rr[\Delta\KK(\rr)]^2\}
\ee
It is easily verified that 
the SCBA (\ref{scba0}) is obtained
by minimizing $S_{\rm eff}$ of Eq. (\ref{seff1}).
This corresponds to the saddle-point
approximation of the effective field theory derived for the
appropriate stochastic Helmholtz equation
\cite{mckane81,john83,maurer04a,schirm06}. The saddle-point
approximation relies on the large prefactor
of $S_{SCBA} \propto 1/\gamma$, i.e.
the SCBA has its validity range for
\begin{itemize}
	\item[($i$)]small disorder $\langle K^2\rangle/K_0^2\ll$ 1;
	\item[($ii$)]Gaussian disorder, Eq. (\ref{gauss}).
\end{itemize}

As in our calculation of the Rayleigh scattering and the
non-analyticity of $D(s)$ the SCBA transforms the microscopic
wave equation with fluctuating elastic coefficient

\be
\big[-z^2 -\nabla\KK(\rr)\nabla \big]u(\rr,z)
\ee
into a macroscopic mean-field wave equation
\be
\big[-z^2 -\KK(z)\nabla^2 \big]u(\rr,z)
\ee
The fluctuations of $\KK(\rr)$, (represented by the variance)
determine the frequency dependence of $\KK(z)$.

It is useful to formulate the SCBA, Eq. (\ref{scba0}) in dimensionless
units (indicated by a hat). We measure velocities in units of $\sqrt{\KK_0}$,
lengths in units of $k_\xi$, and angular frequencies in units of
$k_\xi\sqrt{\KK_0}$. (In the diffusion problem diffusivities
are measured in units of $D_0$ and angular frequencies
in units of $D_0k_\xi^2$.) In these units the SCBA, Eq. (\ref{scba0}) takes
the form
\be\label{scba1}
\widehat\Sigma(\hz)=\frac{\Sigma(z)}{\KK_0}
=3\widehat\gamma \int_0^{1}d\hk \hk^2\frac{\hk^2}{-\hz^2+\hk^2[1-\widehat\Sigma(\hz)]}
\ee
with the dimensionless ``disorder parameter''
\be
\widehat\gamma=\gamma\frac{\nu}{\KK_0^2}=V_c\nu\frac{\langle(\Delta\KK)^2\rangle}{\KK_0^2}
\ee
For $\hz = 0$ Eq. (\ref{scba1}) takes the form
\be
\widehat\Sigma(0)=\frac{\widehat \gamma}{1-\widehat\Sigma(0}
\ee
This quadratic equation has the solution
\be
\widehat\Sigma(0)=\frac{1}{2}\bigg[-1+\sqrt{1-4\widehat\gamma}\bigg]
\ee
We observe that for $\widehat\gamma>\widehat\gamma_c=1/4$
no real solution is obtained. This indicates an instability, 
i.e. for $\widehat\gamma>\widehat\gamma_c$ the SCBA predicts eigenvalues
$\omega^<0$.
This can be rationalized by the fact
that the SCBA is obtained from assuming a Gaussian distribution
(\ref{gauss}). If the width of this distribution exceed a critical
value, there exist too many local regions with negative elastic
coefficients, which then leads to the instability. Within this
model a strong boson peak is obtained if the disorder parameter
$\widehat\gamma$
approaches the critical value $\widehat \gamma_c$.

If we assume that the Debye cutoff $k_D=\sqrt[3]{6\pi^2N/V}$
coincides with $k_\xi$ ($N$ is the number of atoms or molecular
units and $V$ the sample volume),
the density of states is given by
\be\label{dos0}
g(\hat\omega)=\frac{2\hat\omega}{\pi}
3\int_0^1 d \hk \hk^2
\im\left\{\frac{1}{-\hz^2+k^2\big[1-\widehat\Sigma(\hz)\big]}\right\}
\ee

\begin{figure}
\includegraphics[width=.45\textwidth]{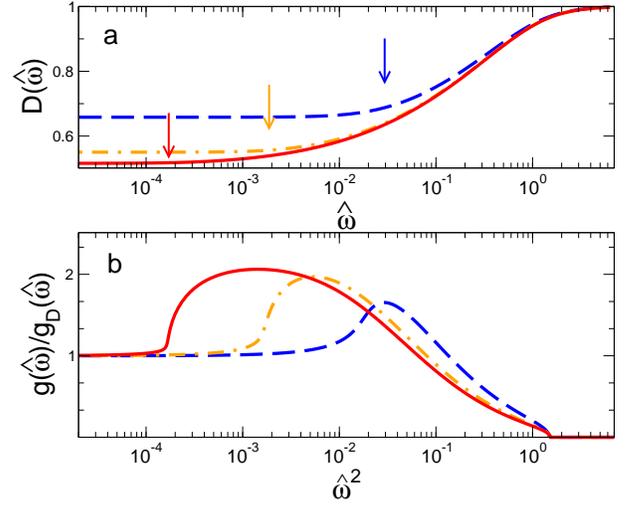}
	\caption{Panel a:
	Frequency-dependent diffusivity, calculated in
	SCBA, Eq. (\ref{scba1}) with disorder parameters
	$(\hat\gamma-\hat\gamma_c)/\hat\gamma_c=$
	10$^{-1}$ (blue dashes),
	10$^{-2}$ (orange dash-dots),
	10$^{-3}$ (red line).\\[.1cm]
	Panel b: Reduced density of states $g(\omega)/g_D(\omega)$
	for the same parameters as in panel a and setting
	$k^\xi=k_D$.
		}\label{picbos}
\end{figure}

In Fig. \ref{picbos} we have plotted the frequency-dependent
diffusivity (panel a) together with the DOS, divided by the Debye
DOS $g_D{\hat\omega}=3\hat\omega^2/\hat\omega_D^3$
(reduced DOS),
with $\hat\omega_D=\sqrt{1-\widehat\Sigma_0}$,
for different values of $\widehat\gamma$ near $\widehat\gamma_c$.
We see that the boson peaks indeed coincide with the
onset of the frequency dependence of $D(\omega)$.

\subsection{Strong disorder and the coherent-potential approximation (CPA)}
It is clear from Fig. \ref{picbos} that the frequency dependence of
the diffusivity (and hence of the AC conductivity), predicted by the SCBA
is rather weak, as compared to the strong frequency dependence of the
conductivity in ionic conducting glasses
or amorphous semiconductors \cite{jonscher77,long82}. This is so, because
the SCBA is restricted to very 
weak disorder. In materials with activated ionic or electronic hopping
conduction, on the other hand, the local diffusivities fluctuate very
strongly, because they depend exponentially on local activation energies
and local tunneling distances \cite{bottger85,efros84}.
\begin{figure}
\includegraphics[width=.45\textwidth,clip=true]{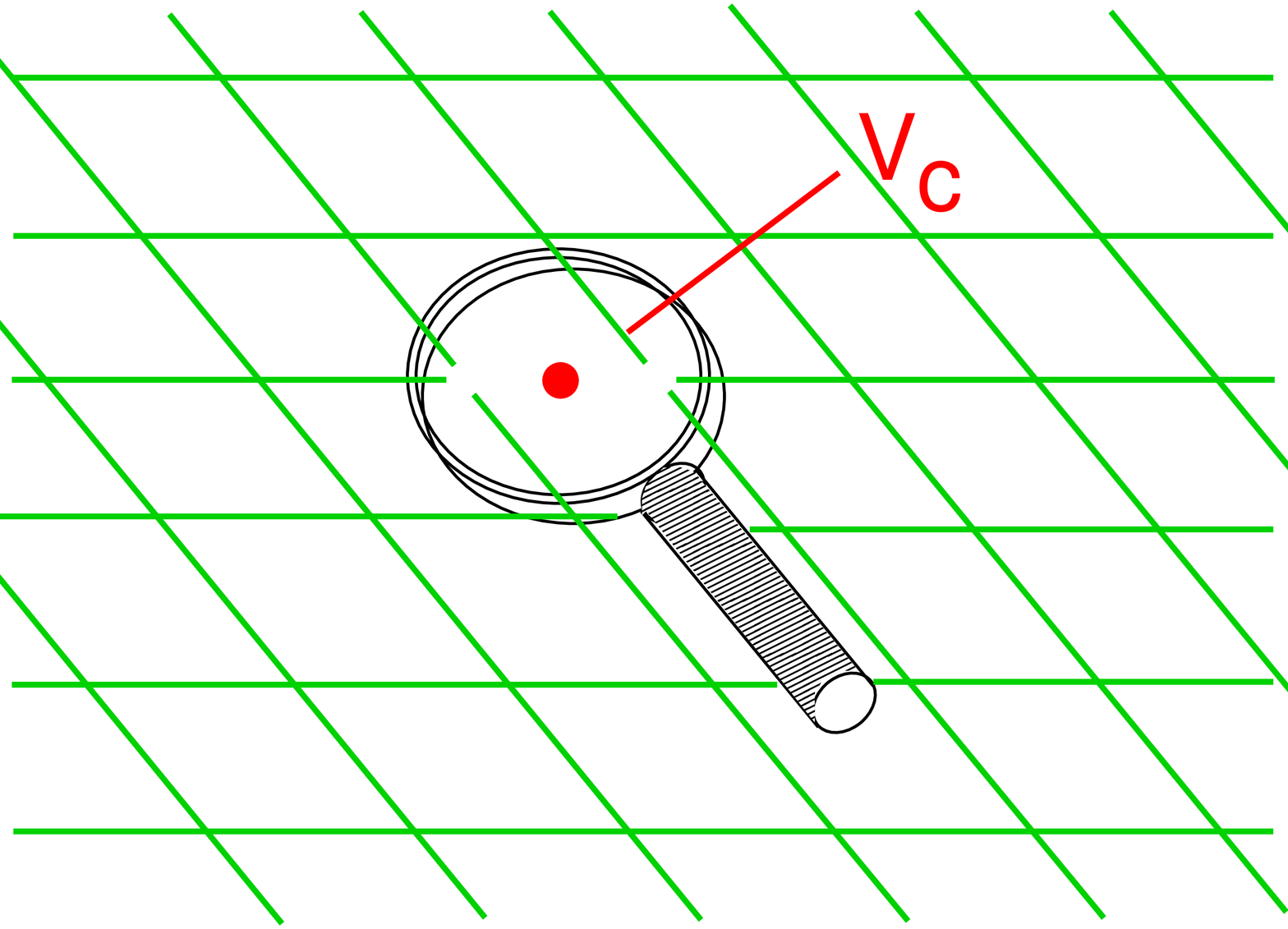}
	\caption{Visualization of the continuum version of the CPA:
	Inside a volume $V_c$ of the effective medium with
	homogeneous elasticity $Q(z)$ the homogeneous one is
	replaced by the fluctuating one $\KK(\rr)$, which gives
	rise to a perturbation $\KK(\rr)-Q(z)$. The CPA postulate
	is to minimize the influence of this perturbation, i.e.
	forcing the averaged $T$ matrix of the perturbation 
	to be equal to zero.
		}\label{lupe4a}
\end{figure}

A rather successful and reliable mean-field theory for strong disorder
is the coherent-potential approximation (CPA). It was widely used
for electronic structure calculations of disordered crystals
\cite{soven67,taylor67,ebert11}, and later to diffusion and
vibrational properties of disordered crystals \cite{webman81,odagaki81,summerfield81,schirm98,te01}. In this lattice version of the CPA one consider a certain
lattice site of the ordered effective medium, in which the potentials
\cite{soven67,taylor67,ebert11},
the
force constants or the diffusivities 
\cite{webman81,odagaki81,summerfield81,schirm98,te01}
are homogeneous (``coherent'')
but frequency-dependent.
At this special site the effective medium is replaced by the real medium,
causing a ``perturbation'' of the effective medium. Enforcing now the
averaged $T$ matrix of this perturbation to vanish gives the self-consistent
CPA equation for the coherent potential.

A version suitable for non-crystalline materials
has been worked out by S. K\"ohler and the present authors using field-theoretical techniques. 
\cite{kohler13}. 
The resulting CPA equation may be pedagogically visualized
in the following way, see Fig. \ref{lupe4a}: 
in a certain region of the effective medium the frequency-dependent
elasticity
$Q(z)$ is replaced by the fluctuating one $\KK(\rr_i)$, where
$\rr_i$ is the mid-point of the region. The CPA postulate takes the form
\be\label{cpa1}
\left\langle
\frac{\KK(\rr_i)-Q(z)}{1+\Lambda(z)[\KK(\rr_i)-Q(z)]}
\right\rangle=0
\ee
The average is over the distribution of the elasticity values
$\KK(\rr_i)\equiv \KK_i$ with distribution density
$P(\KK_i)$. The latter may be taylored to the statistics
of the disordered material at hand.

$\Lambda (z)$ generalizes the effective-medium propagator in the
lattice CPA. Both $\Lambda(z)$ and $Q(z)$
may be obtained 
by minimizing
the following mean-field action

\be\label{seffcpa}
S_{\rm eff}[Q(z),\Lambda(z)]=S_{\rm med}[Q(z)]
+S_{\rm CPA}[Q(z),\Lambda(z)]
\ee
where $S_{\rm med}[Q(z)]$ is given by Eq. (\ref{seff1}) and
the CPA action by
\be\label{seffcpa1}
S_{CPA}[Q(z),\Lambda(z)]=
\frac{V}{V_c}\bigg\langle
\ln
\bigg(
	1+\Lambda(z)[\KK(\rr_i)-Q(z)]
	\bigg)
\bigg\rangle
\ee
The large
parameter of the saddle-point approximation, which validates the
mean-field approximation, is now {\it not} the inverse disorder parameter,
as in the case of the SCBA, but the parameter $V/V_c$. This is the
reason, why the CPA is not restricted to small disorder.
Varying the action with respect to $\Lambda(z)$ we obtain the CPA equation
(\ref{cpa1}),
which can be put into the equivalent forms
\be\label{cpa2}
\left\langle
\frac{1}{1+\Lambda(z)[\KK_i-Q(z)]}
\right\rangle=1
\ee
\be\label{cpa3}
Q(z)=
\left\langle
\frac{\KK_i}{1+\Lambda(z)[\KK_i-Q(z)]}
\right\rangle\, .
\ee
Varying the action with respect to $Q(z)$ gives a relation
for $\Lambda(z)$:
\ba
\sum_\kk\frac{k^2}{-z^2+k^2Q(z)}
&=&
\frac{V}{V_c}\Lambda(z)
\left\langle
\frac{1}{1+\Lambda(z)[\KK_i-Q(z)]}
\right\rangle\nonumber\\
&=&
\frac{V}{V_c}\Lambda(z)\, ,
\ea
where the second line follows from 
Eq. (\ref{cpa2}).
\begin{figure}
\includegraphics[width=.45\textwidth]{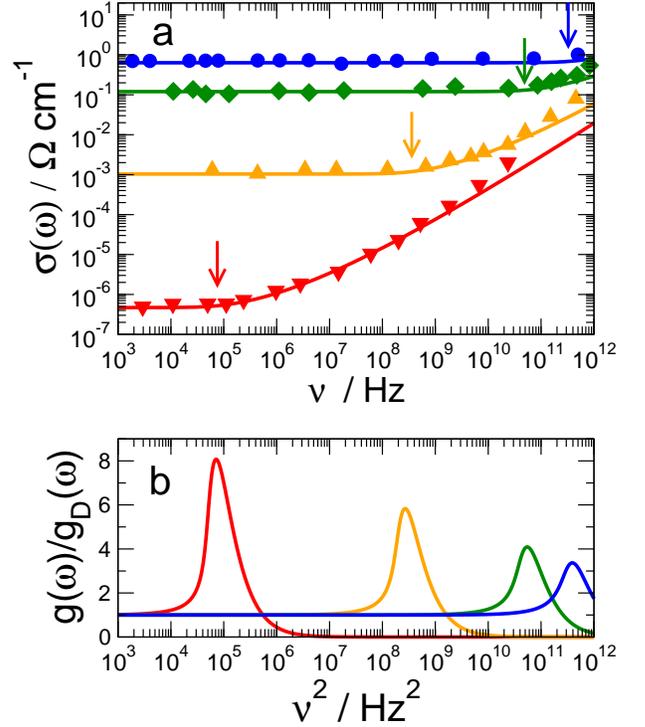}
	\caption{Panel a:
	Full lines:
        Frequency-dependent diffusivity, calculated in
	CPA, Eq. (\ref{cpa4}), with a flat distribution
	of activation energies, Eq. (\ref{flat}). Symbols: AC conductivity data
        of the ionic-conducting glass Sodium Trisilicate, compiled
	by Wong and Angell \cite{wong}.
	with disorder parameters $\widetilde\gamma$
	= 7.92 ($\leftrightarrow T$ = 1673 K);
	= 10.41 ($\leftrightarrow T$ = 1273 K);
	= 17.53 ($\leftrightarrow T$ = 756 K);
	= 29.06 ($\leftrightarrow T$ = 456 K);
	(from top to bottom).\\[.1cm]
	Panel b: Reduced density of states $g(\hat\omega)/g_D(\hat\omega)$
	for the equivalent distribution density (\ref{inverse})
        for the same parameters as in panel a.
		}\label{cpanum}
\end{figure}

It is advantageous to
normalize the $\kk$ integration to 1 
\be
\Lambda(z)=\frac{V_c}{V}
\sum_\kk\frac{k^2}{-z^2+k^2Q(z)}
=p\chi^{\xi}(z)
\ee
with
\be
\chi^\xi(z)=\frac{3}{k_\xi^3}\int_0^{k_\xi}dk k^2
\frac{k^2}{-z^2+k^2Q(z)}
\ee
and
\be
p=\frac{V_ck_\xi^3}{6\pi^2}
\ee
$p$ should be smaller than 1,
and It has been argued in Ref. \cite{kohler13} that one may
identify $p$ with the continuum percolation concentration $p_c$.
One has $\chi^\xi(z\!=\!0)=1/Q(0)$. 

The CPA equation (\ref{cpa3})
now takes the form
\be\label{cpa4}
Q(z)=
\left\langle
\frac{\KK_i}{1+p\chi^\xi(z)[\KK_i-Q(z)]}
\right\rangle\, .
\ee
As in the case of the SCBA the CPA turns the wave equation
with fluctuating elasticity into a mean-field wave equation
with frequency-dependent elastic coefficient $Q(z)$
or a diffusion equation with frequency-dependent diffusivity $D(s)$.

Let us now identify $\KK_i$ with a spatially fluctating
diffusivity $D_i$ and assume that the fluctations are caused
by a fluctuating activation energy
\be
D_i=D_0e^{-E_i/k_BT}
\ee
If we now impose a constant distribution density of activation
energies
\be\label{flat}
P(E_i)\frac{1}{E_c}\theta(E_c-E_i)
\ee
i.e. a flat distribution with cutoff $E_c$. For the diffusivities
$D_i$
(or elasticities $\KK_i$) this transforms to a truncated inverse-power
distribution \cite{kohler13}
\be\label{inverse}
P(\KK_i)=\frac{1}{\mu_1/\mu_2}\frac{1}{\KK_i}\qquad
\mu_1\le\KK_i\le\mu_2
\ee
with $\mu_1=\mu_2e^{-E_c/k_B}$, and we have
\be
\frac{1}{\langle\KK\rangle^2}\langle\KK^2\rangle-\langle\KK\rangle^2)
=\frac{1}{2}E_c/k_BT\equiv\widetilde\gamma
\ee
Low temperatures in the diffusion problem obviously mean strong disorder.

In panel a of Fig. \ref{cpanum} we have plotted AC conductivity data for
the glassy ionic conductor Sodium Trisilicate \cite{wong}. Together with
these data we plot the CPA result with a distribution of activation
energies as given in Eq. (\ref{flat}). In panel b of this figure we
have plotted the reduced DOS of the equivalent vibrational problem
with inverse-power distribution (\ref{inverse}). 
The DOS for the scalar phonon problem is given by
\be
g(\omega)=\frac{2\omega}{\pi}\im\bigg\{
	\frac{1}{N}\sum_{|\kk|\le k_D}
	\frac{1}{-z^2+k^2Q(z)}\bigg\}
\ee
Apart from the perfect
agreement between the conductivity data and the CPA curves we note
that the (rather strong) boson peak precisely mark the crossover
from frequency-independent to frequency-dependent conductivity
of the diffusion problem.
\begin{figure}
\includegraphics[width=.45\textwidth]{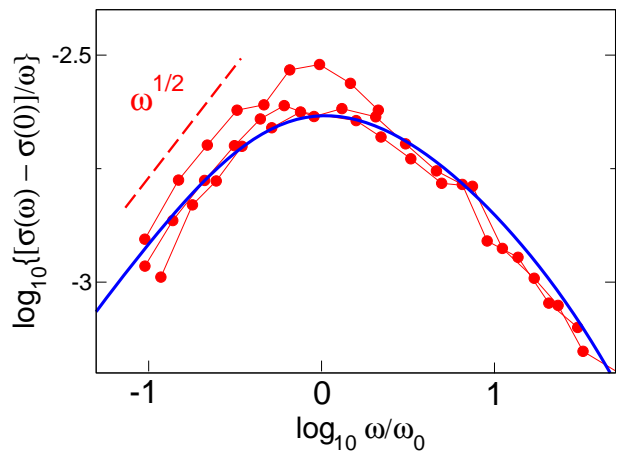}
	\caption{Connected symbols: AC loss function $[\sigma(\omega)-\sigma(0)]/\omega$,
	measured in sputtered amorphous silicon \cite{long88} for three
	different temperatures ($T$ = 51 K, 77 K, 154 K). The scaling frequency
	$\omega_0(T)$ is
	proportional to $\sigma(0,T)$. Full thick line: CPA calculation
	for a constant distribution of activation energies.
		}\label{long2}
\end{figure}
In the vibrational problem this crossover means  a transition from a regime 
with Debye waves and Rayleigh scattering to a non-Debye regime, in which
the vibrational excitations are not waves. These excitations have
been called ``diffusons'' \cite{allen99}, because their intensity
obeys a diffusion equation like light in turbid media
\cite{ishimaru78}. On the other hand, the vibrational excitations in
this regime show the statistical properties of random matrices
\cite{schirm98,matharoo04,beltukov13}. Therefore the frequency regime
above the boson peak may also be called ``random-matrix regime''
\cite{schirm13,schirm14}.

Returning to the diffusion problem, we pointed out that in the
frequency range below the crossover (the frequency-independent regime) the Rayleigh
scattering corresponds to a contribution to the frequency dependent
conductivity with $\Delta \sigma'(\omega)\propto\omega^{3/2}$.
This may be experimentally verified by considering the dielectric
loss function
\be
\epsilon''(\omega)\propto \frac{1}{\omega}\big[
	\sigma'(\omega)-\sigma(0)
	\big]
\ee
In the frequency-independent regime this function {\it increases} with frequency
$\epsilon''\propto\omega^{1/2}$, whereas in the frequency-dependent regime
it {\it decreases} due to the sublinear behavior of $\sigma'(\omega)$.
The maximum of the loss function corresponds to the maximum of
the reduced DOS in the vibrational problem (boson peak).

It has been noted in the literature \cite{dyre00} that the AC conductivity
data taken at different temperatures
show universal behavior, if the data are divided by the DC conductivity
$\sigma(0,T)$ and the freqency by the crossover frequency $\omega_0(T)\propto
\sigma(0,T)$. For the loss function this means that $\epsilon''(\omega/\omega_0)$
should also be the same for different temperatures.

In Fig. \ref{long2} we show the loss function of sputtered amorphous
silicon \cite{long88}
against $\omega/\omega_0$ for three temperatures, together with
the result of the CPA, which predict the scaling and the crossover
from $\epsilon''\propto \omega^{1/2}$ to the decrease with frequency.
We discuss the corresponding scaling of the vibrational DOS
in section III D.

\section{Heterogeneous-elasticity theory}
\subsection{Model}
We now formulate the full heterogeneous-elasticity theory
for vector displacements $\uu(\rr,t)$
The equation of motion for an elastic medium with spatially fluctuating
shear modulus $G(\rr)$\footnote{As locally the translational and
rotational invariance is broken one should in principle work with
the full fourth-rank Hooke tensor $C_{ijkl}(\rr)$ instead of equation
(\ref{eqmo1}). The latter is an approximation to keep the model
tractable.}

\be\label{eqmo1}
\big[\frac{\partial^2}{\partial t^2}
-\nabla\cdot\MM(\rr)\nabla\cdot\,\,
+\nabla\times \GG(\rr)\nabla\times\big]\uu(\rr,\omega)=0
\ee
with the reduced shear modulus $\GG(\rr)=G(\rr)/\rho$
and the reduced longitudinal modulus
\be
\MM(\rr)=M(\rr)/\rho=\KK+\frac{4}{3}\GG(\rr)
\ee
In this formulation the dilatational (bulk) modulus 
$K=\KK/\rho$ is assumed
not to exhibit spatial fluctuations, i.e. the fluctuations 
of the shear modulus $G(\rr)=\GG(\rr)/\rho$
are assumed to
affect the traceless stress and strain tensors
only \cite{kohler13,schirm14}.

Eqs. (\ref{eqmo1}) may be decoupled by introducing longitudinal
and transverse displacements $\uu_L(\rr,t)$ and $\uu_T(\rr,t)$ with
\be\label{longtrans}
\nabla \times\uu_L(\rr,t)=0\qquad\mbox{and}\qquad\nabla\cdot\uu_T(\rr,t)=0
\ee

In the frequency domain we then have

\ba\label{eqmolong}
0&=&\bigg(
-z^2
-\nabla\cdot\MM(\rr)\nabla\cdot\bigg)\uu_L(\rr,z)\nonumber\\
&\equiv&A_L[z,\rr,\GG(\rr)]\uu_L(\rr,z)
\ea
\ba\label{eqmotrans}
0&=&\bigg(-z^2
+\nabla\times \GG(\rr)\nabla\times\bigg)\uu_T(\rr,z)\nonumber\\
&\equiv&A_T[z,\rr,\GG(\rr)]\uu_T(\rr,z)
\ea
\subsection{Self-consistent Born approximation}

As in the case of the scalar model, the SCBA and the CPA,
(see next subsection) serve to calculate the frequency dependence
of the reduced frequency-dependent shear modulus 
\ba\label{trans}
Q(z)&=&G(z)/\rho
=Q'(\omega)-iQ''(\omega)\nonumber\\
&=&v_T(z)^2=\GG_0-\Sigma(z)
\ea
and 
longitudinal modulus 
\ba\label{long}
\MM(z)&=&M(z)/\rho=\KK+\frac{4}{3}Q(z)
=\MM'(\omega)-i\MM''(\omega)\nonumber\\
&=&v_L(z)^2=\KK+\frac{4}{3}[\GG_0-\Sigma(z)]
\ea
which enter
into macroscopic mean-field equations of motion
\be\label{eqmolongz}
0=
-z^2
-\MM(z)\nabla^2
\uu_L(\rr,z)
\equiv A_L[z,\rr,Q(z)]\uu_L(\rr,z)
\ee
\be\label{eqmotransz}
0=
-z^2
-Q(z)\nabla^2
\uu_T(\rr,z)
\equiv A_T[z,\rr,Q(z)]\uu_T(\rr,z)
\ee
The effective action for deriving the SCBA is
\be\label{seffscba}
S_{\rm eff}[\Sigma(z)]=S_{\rm med}[\Sigma(z)]+S_{SCBA}[\Sigma(z)]
\ee
Here $S_{SCBA}[\Sigma(z)]$ is given by Eq. (\ref{seff2})
with
\be
\gamma=V_c\big\langle [\GG(\rr)-\GG_0]^2\big\rangle
\ee
The difference from the scalar model is that we now deal with
3-dimensional vectors. Therefore the trace includes a sum
over the 3 cartesian degrees of freedom. This means that
one has to sum the longitudinal contribution once and that
of the transverse twice. Explicitly $S_{\rm med}[\Sigma(z)]$ takes the form
\ba\label{smedvec}
&&S_{\rm med}[\Sigma(z)]=
\tr\ln\bigg(A\big[z,Q(z)\big]\bigg)\nonumber\\
&&=
\sum_\kk
\ln\bigg(A_L\big[z,\kk,\GG_0-\Sigma(z)\big]\bigg)\nonumber\\
&&+2\sum_\kk
\ln\bigg(A_T\big[z,\kk,\GG_0-\Sigma(z)\big]\bigg)
\ea
If we now vary $S_{\rm eff}$ of Eq. (\ref{seffscba}) with respect
to $\Sigma(z)$, i.e. 
$
\frac{
\partial S_{\rm eff}
}{
\partial \Sigma(z)
}=0$
we get
\be\label{scba2}
\Sigma(z)=\gamma \frac{k_\xi^3}{6\pi^2}
\chi^\xi(z)
\ee
with 
\be\label{susc}
\chi^\xi(z)=
\frac{4}{3}\chi_L^\xi(z)+2\chi_T^\xi(z)
\ee
This weighted susceptibility is given in terms of the local
longitudinal and transverse susceptibilities
\be\label{suscxi}
\chi_{L,T}^\xi(z)=\frac{3}{k_\xi^3}\int_0^{k_\xi}\!\!\!dkk^2
\chi_{L,T}(k,z)
\ee
with the k dependent susceptibilities
\be\label{susck}
\chi_{L,T}(k,z)=k^2\Gg_{L,T}(k,z)=\frac{k^2}{
	-z^2+k^2v_{L,T}(z)^2}
\ee
$\Gg_{L,T}(k,z)$ are the longitudinal and transverse Green's functions%
\footnote{We use sans serif for the Green's functions $\Gg_{L,T}(k,z)$ 
in order
to distinguish them from the shear modulus $G$.}
Eq. (\ref{scba2}) together with Eqs. 
(\ref{trans}),
(\ref{long}),
(\ref{susc}),
(\ref{suscxi}), and
(\ref{susck}) establish the self-consistent vector SCBA equations 
for calculating the frequency-dependent reduced shear modulus
$Q(z)$, and from this the relevant measurable quantities:
\begin{itemize}
	\item Vibrational density of states
\be\label{dos}
g(\omega)=
\frac{2\omega}{3\pi}
\frac{3}{k_D^3}\int_0^{k_D}dk k^2
\bigg(\Gg_L''(k,\omega)+2\Gg_T''(k,\omega)\bigg)
\ee
\item Specific heat
\be
C(T)\propto \int_0^\infty d\omega g(\omega)(\omega/T)^2
\frac{e^{\hbar\omega/k_BT}}{[e^{\hbar\omega/k_BT}-1]^2}
\ee
\item Longitudinal and transverse sound attenuation $\Gamma_{L,T}$
\ba\label{gammal}
\Gamma_L(\omega)&=&\omega M''(\omega)/M'(\omega)\nonumber\\
&&\nonumber\\
\Gamma_T(\omega)&=&\omega G''(\omega)/G'(\omega)\\
&&\nonumber\\
\Leftrightarrow\quad
		v_{L,T}(z)^2&=&\re\big\{v_{L,T}^2\big\}\big[1-i\Gamma_{L,T}/\omega\big]\nonumber
\ea
\item Longitudinal and transverse mean-free paths $\ell_{L,T}$
\be
\frac{1}{\ell_{L,T}(\omega)}=\frac{\Gamma_{L,T}(\omega)}{2v_{L,T}(0)}
\ee
\item Thermal conductivity \cite{schirm06}
\be\label{kappa}
		\kappa(T)\propto \int_0^\infty d\omega \ell_T(\omega)(\omega^4/T^2)
\frac{e^{\hbar\omega/k_BT}}{[e^{\hbar\omega/k_BT}-1]^2}
\ee
\item Coherent inelastic neutron and X-ray scattering intensity
	\be\label{skom}
		S(k,\omega)\propto \underbrace{\frac{1}{1-e^{\hbar\omega/k_BT}}}_{\ds n(\omega)+1}
                \im\big\{\chi_L(k,\omega)\big\}
		\ee
\item Depolaritzed Raman scattering intensity \cite{schmid08,schirm14}
	\be\label{raman}
		I_{VH}(\omega)\propto[n(\omega)+1]
		\im\big\{\chi^{k_p}(\omega)\big\}
		\ee
\end{itemize}
It should be noted that the same susceptibility combination, 
namely $\chi^\xi(\omega)$, which
enters into the SCBA equation (\ref{scba2}), appears also in
the Raman intensity, Eq. (\ref{raman}), albeit
with a cutoff $k_P$ given by the fluctuations of the pockels constants
and not the elastic constants \cite{schmid08,schirm14}. 

It should further
be noted that the Raman intensity is {\it not} given
by the Shuker-Gammon formula 
\mbox{$I_{VH}(\omega)\propto [n(\omega)+1]g(\omega)/\omega$} \cite{shuker71}. If one divides the expression of Eq. (\ref{raman}) for $I_{VH}(\omega)$
by Shuker and Gammon's expression one obtains the frequency-dependent
coupling constant $C(\omega)$, which had been inserted as additional
factor into the Shuker-Gammon formula in order to reconcile
neutron and Raman-scattering results for the vibrational DOS
\cite{jackle81,malinovsky86,viliani95,schmid08,novikov09,schmid09}.

\begin{figure}
\centerline{\includegraphics[width=8cm,clip=true]{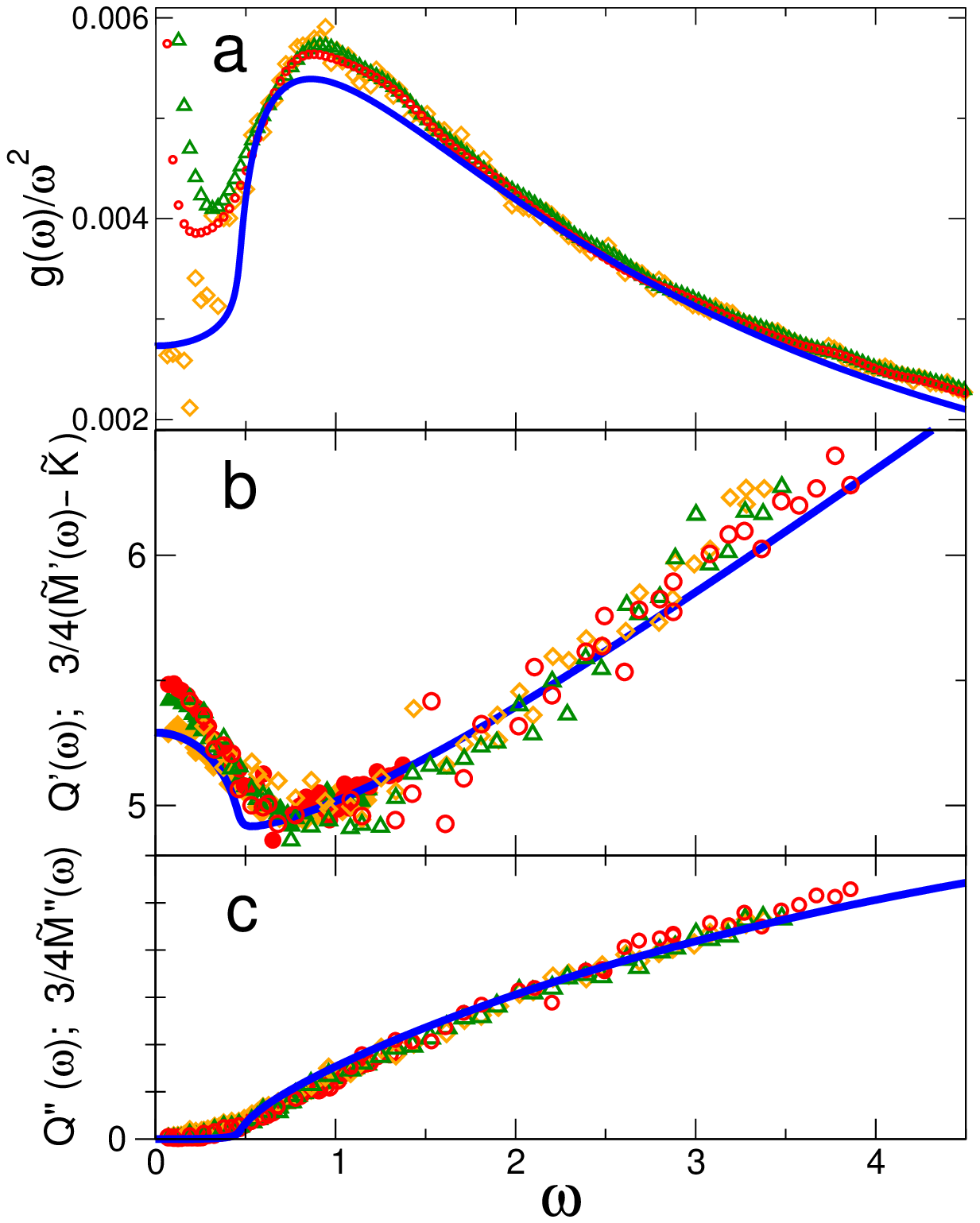}}
\caption{
Comparison of results of a soft-sphere molecular-dynamics
simulation (symbols) with the prediction of heterogeneous-elasticity
theory in self-consistent Born approximation (SCBA). We show the
real parts (upper panel) and imaginary
parts (lower panel)
of the frequency-dependent
shear modulus $G(\Omega_T)$ and the
quantitity $3/4(M(\Omega_L)-\tilde K)$ 
for three 
temperatures 
($5\cdot 10^{-3}$,
$5\cdot 10^{-4}$,
$5\cdot 10^{-5}$, in Lennard-Jones units)
with $K=30.4$. The SCBA parameters are
$\gamma-\gamma_c=0.08$ and $K/G_0 =3.166$;
from \cite{marruzzo13}.
} \label{figg6}
\end{figure}
\subsection{General features of the disorder-induced vibrational
anomalies: Comparison of the SCBA version of heterogeneous-elasticity
theory with a simulation}

As stated in the introduction, the disorder-induced vibrational anomalies
of glasses and other disordered solids feature three phenomena, which
are related to each other
\begin{itemize}
	\item[($i$)] 
		The cross-over from Debye- to non-Debye behavior
		of the vibrational DOS, leading to a maximum
		in the reduced DOS $g(\omega)/\omega^2$, the boson peak;
	\item[($ii$)] a pronounced dip in the real part of the
		elastic coefficients (and their square-root, the
		frequency-dependent velocities) near the boson-peak frequency;
	\item[($iii$)] 
		a strong increase of the sound attenuation below the
		boson-peak frequency
		(Rayleigh scattering), which enters into the imaginary
		parts of the elastic coefficients via Eq. (\ref{gammal}).
\end{itemize}

These three anomalies are displayed in Fig. \ref{figg6}, in which
the results of a molecular-dynamic simulation is compared with the
prediction of heterogeneous-elasticity theory, solved in SCBA
\cite{marruzzo13}.
In this simulation a binary soft-sphere potential (i.e. a
Kob-Andersen-type \cite{kobandersen94}
binary Lennard-Jones potential without the attractive part)
was taken for ten million particles. Such a potential mimics
a metallic glass.
The longitudinal and transverse frequency-dependent moduli were
obtained from determining the longitudinal and transverse current
correlation functions
\be
C_{L,T}(k,\omega)=\frac{K_BT\omega}{\pi m}\Gg''(k,\omega)
\ee
where $m$ is the paticles' mass.
The simulation was run for three
very different temperatures deep in the glassy state (see the figure
caption). The very large particle number made it possible to avoid
finite-size effect in the boson-peak frequency range. The data for the
complex longitudinal modulus was converted to 
$Q(z)=G(z)/\rho$ via the inverse
of Eq. (\ref{long})
\be
Q(z)=\frac{3}{4}[\MM(z)-\KK]
\ee
in order to compare the data with the measured $Q(z)$. It is seen
in the figure that -- as the longitudinal and trasverse
data ly on top of each other --
the bulk modulus $\KK$ indeed does not 
appreciably depend on frequency and correspondingly has a negligible
imaginary part.

It is the strength of the present theory that it easily allows to 
explain how the three anomalies are related to each other. It was 
pointed out by Schirmacher et al. 2007 \cite{schirm07} that one can deduce
from Eq. (\ref{dos}) and using the 3rd line of (\ref{gammal}) a
relation between the DOS and $\Gamma(\omega)$:
\be\label{dos1}
g(\omega)-g_D(\omega)\propto \Gamma(\omega)\propto \omega\Sigma''(\omega)
\ee
This means that the disorder-induced frequency dependence of the
elastic coefficient, controlled by $\Sigma(\omega)$ is responsible
for the boson peak. This can already understood using lowest-order
perturbation theory, which leads to Rayleigh scattering.

On the other hand, the real parts and the imaginary parts of
the frequency-dependent moduli are related to those of
the response functions $\chi(z)=\chi'(\omega)-i\chi''(\omega)$.
These have, due to the causality requirement (the answer must occur
at a time later than the question),
a one-to-one correspondence by the Kramers-Kronig relation
\cite{jackson}
\be
\chi'(\omega)=\frac{1}{\pi}P\int_0^\infty d\bar\omega^2\frac{\chi''(\bar\omega)}
{{\bar\omega}^2-\omega^3}
\ee
This is most easily visualized by looking at the simulated data
of the real and imaginary parts of $Q(z)=\GG_0-\Sigma(z)$, displayed
in panels b and c. Usually the real part of an analytic function, such
as $\chi^\xi(z)$, has a maximum near the bottom of the spectrum
${[\chi^\xi]}''(\omega)$ 
and a minimum near the top. Now, because 
not $Q(z)$ but $\Sigma(z)$
is proportional to $\chi^\xi(z)$ via the SCBA relation (\ref{scba2}),
$Q'(\omega)$ displays a {\it minimum} near the begin of the 
``disorder spectrum'', or ``random-matrix spectrum''
${[\chi^\xi]}''(\omega)$,
obviously
marked by the boson peak, displayed in panel a.

In the boson-peak regime and above, the data for very different temperatures
collapse, which proves that the anomalies must be of harmonic origin.
This is not so in the very low frequency regime, where anharmonic effects
become visible \cite{marruzzo13a,ferrante13}.

\subsection{Coherent-Potential Approximation (CPA)}
The vector CPA may be derived in the same way as the vector SCBA
from the effective action
\be\label{seffcpa2}
S_{\rm eff}[Q(z)]=S_{\rm med}[Q(z)]+S_{CPA}[Q(z)]
\ee
where $S_{\rm med}[Q(z)]$ is given by Eq. (\ref{smedvec}) 
of the vector SCBA and $S_{CPA}[Q(z)]$ by Eq. (\ref{seffcpa1}) of the
scalar CPA. The vector CPA equation
takes the same form as the scalar one:
\be\label{cpa4}
Q(z)=
\left\langle
\frac{\GG_i}{1+p\chi^\xi(z)[\GG_i-Q(z)]}
\right\rangle\, .
\ee
but now with fluctuating shear moduli $\GG_i$ and the vector
version of $\chi^\xi(z)$, Eq. (\ref{susc}). 

As in the scalar model the
CPA has several advantages compared to the SCBA:

\begin{itemize}
\item[--] one can treat arbitrary distributions $P(G)$;
\item[--] one is not restricted to weak disorder;
\item[--] one needs not (but can) take negative values of $G$
into account.
\end{itemize}
\begin{figure}
\begin{center}
\vspace{1ex}\includegraphics[width=.45\textwidth]{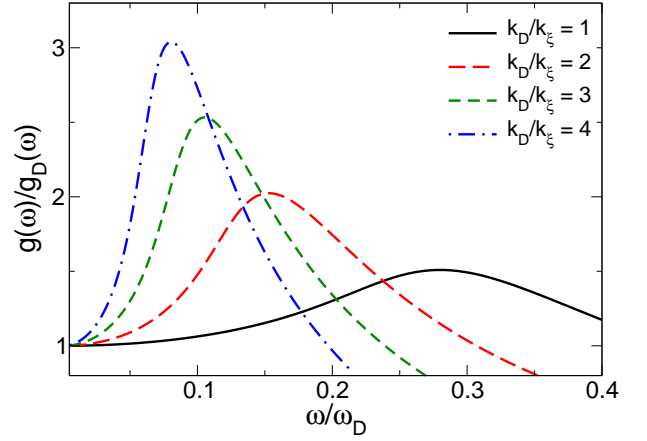}
\caption{\label{fig2} Reduced density of states
$g(\omega)/g_D(\omega)$ vs. the rescaled frequency
$(\omega/\omega_D)(k_D/k_xi)=\omega/v_Dk_\xi$ for
different values of the ratio $k_\xi/k_D$.
The other parameters are $\gamma/G_0^2=1$, $G_{\rm min}=0$ .}
\end{center}
\end{figure}
For the calculations presented in Figs. \ref{fig2} to \ref{fig4}
we used the CPA equation (\ref{cpa4}), together with (\ref{dos})
with a truncated Gaussian distribution of shear moduli of the form
\be
P(G)=P_0\,\theta(G-G_{\rm min})\,e^{-(G-G_0)^2/2\gamma}
\label{heavi}
\ee
where $\theta(x)$ is the Heaviside step function 
and $G_{\rm min}$ is the lower cutoff.
In these calculations we used the {\it renormalized} 
value of $G$ (i.e. the self-consistently calculated one) for
evaluating the
Debye
frequency $\omega_D$ and Debye DOS $g_D(\omega)$
in terms of the longitudinal and transverse
sound velocities 
\mbox{$v_L^2=K+\frac{4}{3}Q(0)$,}
$v_T^2=Q(0)$
\be
\omega_D=k_D\bigg[\frac{1}{3}\left(\frac{1}{v_L^3}+\frac{2}{v_T^3}
\right)\bigg]^{-1/3}
\ee
\be
g_D(\omega)=3\omega^2/\omega_D^3
\ee
For the bulk modulus of the calculations we used
the value $K=3.3G_0$ and for the 
cutoff parameter the value $\widetilde\nu=1$, which implies
$k_D/k_\xi=\sqrt[3]{3}\xi/a$, where $a=\sqrt[3]{V/N}$ is the
mean intermolecular distance.
The distribution of shear moduli (\ref{heavi})
involves three parameters $G_0$, $\gamma$ and $G_{\rm min}$.
Because $G_0$ is used to
fix the elastic-constant scale there remain three adjustable parameters
to fix the {\it state of elastic disorder} of the material, namely
$k_D/k_\xi\sim \xi/a$, $\gamma$ and $G_{\rm min}$. The latter 
(which we used with negative values or equals zero) specify
the amount of regions with negative shear modulus (soft regions)
in the material.
As can be seen from Figs. \ref{fig2} to \ref{fig4}
increasing $\xi$ and $|G_{\rm min}|$ enhances the BP
{\it and} shifts its position to lower frequencies, whereas
increasing $\gamma$ just leads to an enhancement, while keeping
the BP position constant.
It has been pointed out in the literature
\cite{duval90,elliott92,sokolov92,hong11}
that the position of the boson peak in relation to the
Debye frequency correlates with the inverse correlation
length of density and elasticity fluctuations.

\begin{figure}
\begin{center}
\vspace{1ex}\includegraphics[width=.45\textwidth]{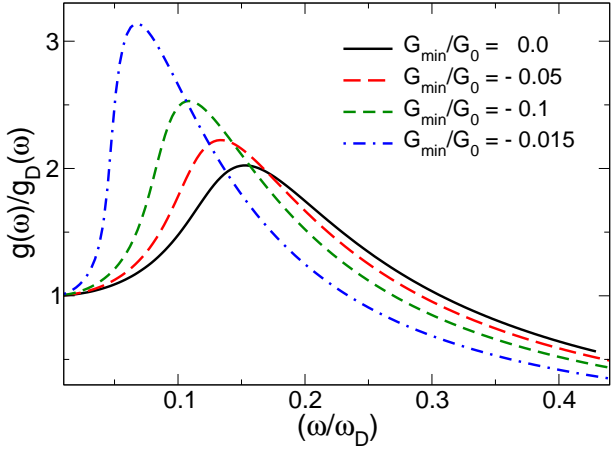}
\caption{\label{fig3} Reduced density of states
$g(\omega)/g_D(\omega)$ vs. the rescaled frequency
$(\omega/\omega_D)$ for
different values of the lower cutoff $G_{\rm min}$ of the
Gaussian distribution $P(G)$. The other parameters are 
$\gamma=/G_0^2=1.$, $k_D/k_\xi=2$.}
\end{center}
\end{figure}

Let us discuss our findings further in terms of measured vibrational
spectra of materials, in which
an external parameter (temperature, pressure or the amount
of polymerization) is changed. If the Debye frequency (depending
on the moduli $K$ and $G$) is changed, this leads to a modification
of the spectrum, which has been called {\it elastic-medium transformation}.
This transformation is taken care of, if the DOS is represented
in a normalized way, as is the case in Figs. \ref{fig2} to \ref{fig2}.
A number of boson-peak data, if normalized in this way, lead to
a universal curve, i.e. all data points fall onto the same
curve if replotted, taking the elastic transformation into
account 
\cite{baldi09,ruta10,chumakov11,chumakov14,monaco06,caponi09}.
Other investigation reveal a {\it deviation} from this
scaling
\cite{zanatta10,zanatta11,hong08,caponi07,ruffle10,niss07,orsingher10,corezzi13,ramos14}.
In terms of our model calculations this means, if the state of disorder
is not changed, but just the value of the mean elastic constants or
the density, which go into the Debye frequency,
this corresponds to elastic-transformation scaling. In the
other cases obviously the state of disorder is changed by changing the
external conditions.

A very interesting case in which the elastic transformation scaling does
not hold has been reported recently: the case of prehistoric amber.
\cite{ramos14} measured the temperature dependence of the specific
heat of the hyperaged and rejuvenated material. The height of the
boson peak, taken from a $C(T)/T^3$ curve is by 22 \% lower in the hyperaged
material, compared with the rejuvenated one. An elastic transformation using
the change in the Debye frequency determined by the authors would only
lead to a difference by 7.4 \%.

\begin{figure}
\begin{center}
\vspace{1ex}\includegraphics[width=.45\textwidth]{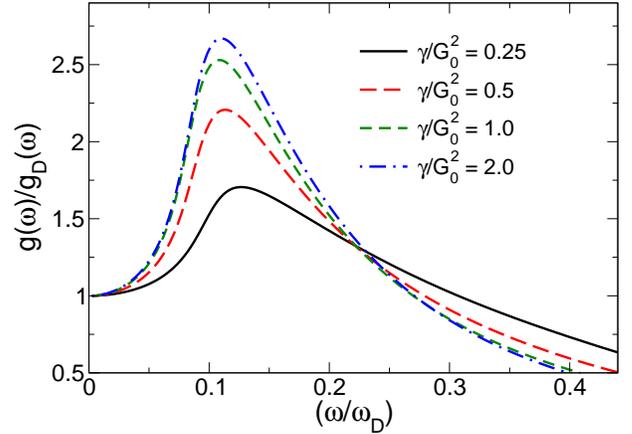}
\caption{\label{fig4} Reduced density of states
$g(\omega)/g_D(\omega)$ vs. the rescaled frequency
$(\omega/\omega_D)$  
for
different values of the  width parameter $\gamma$. The other
parameters are $G_{\rm min}/G_0=-0.1$ and $k_D/k_\xi=2$.}
\end{center}
\end{figure}
\section{Discussion and conclusions}
Reading the text of the present article, one could be convinced that
the boson peak in glasses and the associate vibrational anomalies
can be satisfactory explained by the presence of the structural disorder,
leading to spatial fluctuations of elastic coefficient, in particular
the shear modulus. However, in the community working experimentally
and theoretically on the vibrational
properties of disordered and complex condensed matter there is no agreement
about this. Whereas many authors agree, that the boson-peak-anomalies in
glasses are due to the structural disorder, others maintain that
this is not so and that the anomalies may be explained by 
conventional crystalline solid-state theory. Crystalline phonon
theory is based on the phonon dispersions \cite{mar71}, reflecting the crystal
symmetry group and anharmonic interactions, leading to renormalization
and viscous damping \cite{gotze74,horner74,scheipers96}.

As mentioned in the introduction, Chumakov et al. \cite{chumakov11,chumakov14}
argue that the boson peak is ``identical''
\cite{chumakov11} to a washed-out van-Hove singularity, i.e.
just the result of the bending-over of the lowest (transverse) phonon dispersion
near the first Brillouin zone. In fact, there is experimental evidence
for crystal-like features in the spectrum of glasses
and even liquids \cite{giordano10,baldi13}. These
are caused by the short-range order, which is documented in the static structure
factor $S(k)$ of glasses. The peak position $k_0$
of $S(k)$ corresponds to the diameter of the first Brillouin zone of a crystal.
Correspondingly, $k_0/2$ corresponds to the radius of the Brillouin zone, where
the phonon dispersions become horizontal and cause van-Hove singularities
in the crystalline DOS. 

In their inelastic X-Ray study of polycrystalline
$\alpha$-Quartz and glassy SiO$_2$, densified to match the crystalline
density, accompanied with a numerical lattice-dynamics 
calculation of the crystalline
phonon dispersions, Baldi et al. \cite{baldi13} find a crossover of the sound
attenuation of the glass from a Rayleigh $\Gamma\propto\omega^4$ behaviour
to a quadratic one at $\hbar\omega_c \sim$ 9 meV, where the boson peak of
densified silica has been reported \cite{chumakov14}. At the same frequency
the van-Hove singularity of the polycrystal is located. On the other hand,
in the polycrystal, instead of Rayleigh scattering there is a linear
behavior.
Above $\omega_c$ the two attenuation coefficients match. 

The interpretation of
the authors is that below $\omega_c$ the glassy attenuation is due to
elastic heterogeneities, above $\omega_c$ the spectrum is essentially the same of
that of the polycrystal. These findings are
certainly at variance with the claim
that the boson-peak-type scenario leading to the
$\Gamma\propto\omega^4\rightarrow\omega^2$
crossover would have something to do with the bending-over of the
transverse dispersion of the crystal. Obviously in this material
the boson-peak frequency and the van-Hove-singularity are the same.
This might also be the case for the other examples found by
the authors of refs. \cite{chumakov11,chumakov14}.

As indicated in the introduction, in the investigation of a two-dimensional
macroscopic model glass the co-existence of crystal-like dispersions
with the boson peak was observed \cite{wang18}, but the frequencies of the
boson peak and that of
the transverse van-Hove singularity are quite different. The boson peak was shown
to show all salient features which identify its origin arising from the 
disorder. The conclusion is that the boson-peak, namely the disorder-induced
peak in the reduced DOS is not ``identical'' with a 
washed-out van-Hove singularity. In
some glasses their frequencies are just very near to each other.

A recent interpretation of boson-peak related damping
phenomena claims that the boson peak would be a ``universal phenomenon''
both in crystals and in glasses \cite{zaccone19}. 

The ``universality'' of the boson peak in solids (both perfectly ordered
crystals and glasses), if there is one, suggested by these authors, may be
identified with the fact that the boson peak is linked with a
Ioffe-Regel type of crossover from ballistic phonon propagation to a
scattering-dominated regime where the exctiation is quasi-localized.
This is where the "universality" ends, because the scattering mechanism
is clearly different in glasses and perfectly ordered crystals. In the
former case it is due to (harmonic) disorder, as predicted by
heterogeneous-elasticity theory,
whereas in the latter case of ordered crystals it is due to
anharmonicity.

In conclusion, the boson-peak related vibrational anomalies in glasses can consistently
be explained by the presence of spatial fluctuations of elastic coefficients
(elastic heterogeneity)
with the help of the SCBA and CPA. 

\end{document}